\documentclass[%
 reprint,
 amsmath,amssymb,
 aps,
]{revtex4-1}

\usepackage{graphicx}
\usepackage{dcolumn}
\usepackage{bm}%

\usepackage{braket}
\usepackage{amsthm}

\usepackage{color}
\usepackage{verbatim}

\begin{document}

\title{Dependence of the Quantum Speed Limit on System Size and Control Complexity}

\author{Juneseo Lee}

\author{Christian Arenz} 
\author{Herschel Rabitz}
\author{Benjamin Russell}
\affiliation{
 Chemistry Department, Princeton University\\
}

\date{\today}

\begin{abstract}
We extend the work in New J. Phys. \textbf{19}, 103015 (2017) by deriving a lower bound for the minimum time necessary to implement a unitary transformation on a generic, closed quantum system with an arbitrary number of classical control fields.
This bound is explicitly analyzed for a specific N-level system similar to those used to represent simple models of an atom, or the first excitation sector of a Heisenberg spin chain, both of which are of interest in quantum control for quantum computation.
Specifically, it is shown that the resultant bound depends on the dimension of the system, and on the number of controls used to implement a specific target unitary operation.
The value of the bound determined numerically, and an estimate of the true minimum gate time are systematically compared for a range of system dimension and number of controls; special attention is drawn to the relationship between these two variables.
It is seen that the bound captures the scaling of the minimum time well for the systems studied, and quantitatively is correct in the order of magnitude.
\end{abstract}

\maketitle

\section{\label{sec:level1}Introduction}
Quantum speed limits characterize the maximum speed a quantum system can evolve towards some specific target state or propagator matrix.
For instance, one type of quantum speed limit especially relevant to quantum computation \cite{PhysRevA.75.042308} characterizes the maximum speed at which a quantum system can evolve towards some target unitary propagator.
During the last decades, quantum speed limits have been developed for closed \cite{Mandelstam1991, MLB} as well as for open quantum systems \cite{PhysRevLett.110.050403, PhysRevLett.111.010402} for a wide range of quantum processes using a variety of proof strategies (for a range of examples see \cite{PhysRevA.95.042314, PhysRevA.90.012303, 1751-8121-48-11-115303, Zhang2014,PhysRevLett.103.160502}).
Furthermore, there is a direct connection between the maximal attainable precision in quantum metrology and the maximum speed at which a quantum system can evolve \cite{SpeedLimitMetrol}.
For current reviews regarding the quantum speed limits we refer to \cite{SpeedLimitsRev,Frey2016}.

The quantum speed limit is of particular importance for the practical implementation of quantum information protocols, since it characterizes their feasibility in terms of time and energy scales.
However, the determination of the relevant time scales to run a quantum algorithm described by a sequence of unitary gates is still a challenging and fundamental problem.
Some progress has recently been made towards characterizing the qubit systems for which unitary operations can be implemented efficiently with adequate control resources \cite{Lloyd, HerschMe}.
However, other than for simple idealized low dimensional systems \cite{OptimalControlSpeedLimit2, ExactCalc1, ExactCalc2, ExactCalc3}, a tight and tractable speed limit formula is still missing for implementing unitary gates depending on the available controls and the system size. 

Typically, unitary gates are implemented using some classical time dependent fields which control the evolution of the system described by some fixed Hamiltonian $H_{0}$ so that the total Hamiltonian becomes time dependent.
Within the framework of quantum control theory \cite{ControlRev1,ControlRev2,BookDalessandro}, the goal is then to design suitably shaped pulses in order to implement a desired gate.
Clearly, in order to be practically applicable, the length of the control pulses, henceforth referred to as the \emph{minimum gate time} $T$, should not exceed a certain time scale on which other effects (such as decoherence) can often no longer be neglected.
Moreover, for time dependent systems, surprisingly little is known about a lower bound on the minimum time to implement a unitary gate.
Furthermore, to the authors best knowledge, no algorithm exists which has been proven to converge to time optimal controls for quantum control systems.
Although some progress has recently been made for a system with a single control field \cite{Us}, for generic quantum control systems a quantum speed limit formula is still missing for the implementation of a gate.
The bound on the minimum gate time developed in \cite{Us} depends on the nature of the gate, the norm of the free evolution Hamiltonian $\Vert H_{0}\Vert^{-1}$, and the highest permitted control field amplitude.
We note that the speed limit captured therein persists with nonzero value in scenarios where the control amplitude is unbounded.
However, the central result of \cite{Us} does not scale with the dimension of the system being considered, nor does it depend on the number of controls being used to implement a gate, as only systems with a single control are considered.
Since numerical simulations suggest that the minimum gate time increases when the dimension of the system is increased \cite{Carenzspinstar, Nori, Daniel}, the bound in \cite{Us} becomes a poor estimate for $T$ for high dimensional systems.
For unconstrained control fields a more accurate lower bound on $T$ should therefore depend on the goal gate $G$, the Hilbert space dimension of the quantum system being considered, the number of controls being used to implement $G$, and the nature of the control coupling captured in the control Hamiltonian matrices.

Herein we extend the work in \cite{Us}, deriving a lower bound for the minimum gate time $T$ for a generic quantum control system consisting of an arbitrary number of controls (see Eq. \eqref{eq:lowerboundfin}) and present a system for which the bound captures the system dimension dependence of the quantum speed limit.
Specifically, for a system consisting of $N$ levels, we show that the obtained bound scales at least as $\mathcal{O}(\sqrt{N})$.
Furthermore, we numerically investigate the tightness of the derived bound and study the interplay between the number of controls and the system size.
We show that increasing the number of levels has a qualitatively and quantitatively similar effect on the minimum gate time as decreasing the number of controls to implement a specific SWAP gate.

\section{\label{sec:level1} Derivation of the bound}
We consider the following control system,
\begin{eqnarray}
\label{eq:controlsystem}
\dot{U}(t)=-iH(t)U(t),~~~~U(0)=\openone,	
\end{eqnarray}
on the unitary group $\text{U}(d)$ consisting of unitary $d\times d$ matrices.
Before we investigate the dependence of the minimum time on the number of system controls, we first tighten the bound obtained in \cite{Us} for a single control field
\begin{align}
\label{eq:controlsystem1}
H(t)=H_{0}+f(t)H_{c},	
\end{align}
where we refer to $H_{0}$ and $H_{c}$ as the drift and the control Hamiltonian, respectively and $f(t)$ is the corresponding control field.
We remark that we assume here that the control field enters in a bilinear way \cite{Elliot}. At the final time $T$, the control field implements the goal unitary transformation $G\in \text{U}(d)$ so that $U(T)=G$.
In order to tighten the bound in \cite{Us} we define a new control system by conjugating the Hamiltonian \eqref{eq:controlsystem1} by arbitrary $V\in \text{U}(d)$ such that $[V,H_{c}]=0$.
The corresponding (conjugated) Hamiltonian $\tilde{H}(t)=V^{\dagger}H(t)V$ then reads 
 \begin{align}
 \tilde{H}(t)=V^{\dagger}H_{0}V+ f(t)H_{c},	
 \end{align}
 noting that $\tilde{U}(T)=V^{\dagger} G V$.
Following the approach in \cite{Us,Nielsenbound} we use $\Vert U(T)-\tilde{U}(T)\Vert\leq \int_{0}^{T}\Vert H(t)-\tilde{H}(t)\Vert dt$, which is valid for any unitarily invariant norm, to obtain 
 \begin{align}
 \label{eq:boundfirst}
 T&\geq \frac{\Vert G-V^{\dagger}GV\Vert}{\Vert H_{0}-V^{\dagger}H_{0}V\Vert },\nonumber \\
 &=\frac{\Vert [G,V] \Vert }{\Vert [H_{0},V] \Vert }.
 \end{align} 
The above lower bound on the time $T$ to implement a goal operation $G$ holds for all $V$ that commute with the control Hamiltonian.
By introducing the stabilizer $\text{Stab}(x)=\{U\in \text{U}(d)\,|\,U^{\dagger}xU=x\}$ for some $x\in\mathfrak{u}(d)$ we obtain the tightest bound by taking the maximum, i.e.,
 \begin{align}
 T\geq \max_{V\in\text{Stab}(iH_{c})}\frac{\Vert[G,V] \Vert }{\Vert [H_{0},V]\Vert }.
 \end{align}
We now turn to deriving a lower bound for the minimum gate time for a control systems with $M$ controls described by the set of control Hamiltonians $\{H_{k}\}_{k=1}^{M}$ so that
\begin{align}
 H(t)=H_{0}+\sum_{k=1}^{n}f_{k}(t)H_{k}.	
\end{align}
Consider any $V\in \bigcap_{k} \text{Stab}(iH_k)$ so that the conjugated Hamiltonian is given by 
\begin{align}
\tilde{H}(t)=V^{\dagger}H_{0}V+\sum_{k=1}^{n}f_{k}(t)H_{k}.	
\end{align}
In this case the lower bound \eqref{eq:boundfirst} still holds for all $V$ in the intersection of the stabilizers of the considered controls.
Thus, the tightest lower bound is obtained by maximizing over $V$, i.e., 
\begin{align}
\label{eq:lowerboundfin}
T\geq \max_{V\in \bigcap_{k} \text{Stab}(iH_k)} 	\frac{\Vert [G,V] \Vert }{\Vert [H_{0},V] \Vert }. 
\end{align}
We remark here that the dimension of the intersection of the stabilizers decreases when more control Hamiltonians are present.
Thus, this gives us a hint that more control decreases the speed limit \eqref{eq:lowerboundfin}, as a smaller set of $V$ is maximized over, which yields a lower maximum bound.
However, the maximization in \eqref{eq:lowerboundfin} is not trivial, despite being very much more tractable than finding shaped control fields which minimize the time to implement a given gate.

In order to analyze the tightness of the obtained bound we proceed by analyzing a specific model (sec. \ref{sec:numerical}).
In particular, we show that \eqref{eq:lowerboundfin} is at least proportional to $\sqrt{d}$.
Using the same model, we further numerically carry out the maximzation in \eqref{eq:lowerboundfin} to analyze the dependence on $d$ and on the number of controls in more detail.
The tightness of the bound is then finally studied by comparing the results with a gradient-based search \cite{Grape, DYNAMOpaper, GateFid2, exactgradient} for the control field to implement a specific $G$ for different gate times $T$.

\section{\label{sec:numerical} N-Level system}
We consider a N-level system described by the Hamiltonian 
\begin{align}
\label{eq:nleveldrift}
h	=\sum_{j=1}^{N-1} \left(\ket{j}\bra{j+1}+\ket{j+1}\bra{j} \right),
\end{align}
such that the normalized drift Hamiltonian reads $H_{0}=\frac{h}{\Vert h \Vert }$, where any, not necessarily unitarily invariant, matrix norm is used.
However, throughout this section, we use the Hilbert Schmidt norm ($\Vert A \Vert=\sqrt{\text{tr}\{A^{\dagger}A\}}$) for which one has $\Vert h\Vert =\sqrt{2(N-1)}$.
Control is exerted through the set of projectors $\{P_{j}\}$, i.e., $H_{j}=\ket{j}\bra{j}$. 
These control Hamiltonians physically correspond to a change in energy of the levels of the system induced by the driving of the control field $f_{j}$.
The Hamiltonian \eqref{eq:nleveldrift} represents the first excitation sector of a spin chain with a nearest neighbour isotropic Heisenberg interaction \cite{Daniel,FullControlSingleAcc}, and it is well known that a single control $P_{1}$ is enough to generate a fully controllable system \cite{FullControlSingleAcc}.
As a goal gate $G$ we take the SWAP operation between the first and the N-th level given by 
\begin{align}
\label{eq:Nlevelgoal}
G=\exp\left(-i\frac{\pi}{2}\left(\ket{1}\bra{N}+\ket{N}\bra{1}\right)\right).
\end{align}
We remark here that the N-level system has recently been analyzed in great detail in \cite{DanielNL} for the case when one control is present and the control field is chosen randomly.
Based on unitary q-designs it was shown that the time to  implement a generic goal gate operation scales as $\mathcal O(N^{3})$.
    
\subsection{Analytical assessment}
In order to get some intuition about the derived lower bound we begin by analyzing \eqref{eq:lowerboundfin} for a specific $V\in \bigcap_{j}\text{Stab}(iP_{j})$.
We take 
\begin{align}
\label{eq:choiceV}
V=\openone-2P_{1},
\end{align}
so that the transformed drift Hamiltonian reads $V^{\dagger}H_{0}V^{\dagger}=H_{0}-2(\ket{1}\bra{2}+\ket{2}\bra{1})$ and the rotated goal gate operation is given by $V^{\dagger}GV=G^{\dagger}$. Since $\Vert G-G^{\dagger}\Vert=\sqrt{8}$ and $\Vert 2(\ket{1}\bra{2}+\ket{2}\bra{1})\Vert=2\sqrt{2}$ we find 
\begin{align}
\label{eq:lowerboundanaNlevel}
T_{\text{SWAP}}&\geq \Vert h\Vert= \sqrt{2\left(N-1\right)}.
\end{align}
Thus, for the (normalized) N-level system given by \eqref{eq:nleveldrift} the minimum gate time for implementing a SWAP operation between the first and the Nth level scales at least as $\propto \sqrt{d}$.
We remark here that the system size dependent behavior of the obtained lower bound is a significant improvement over known quantum speed limits, since they do not depend on the dimension of the Hilbert space of the system being considered.
For instance, the bound in \cite{Us} scales as $\propto\Vert H_{0}\Vert ^{-1}$ yielding for the N-level system and $G$ given above that the lower bound is $T\geq 2$ for all $N$. 
Thus, the bound in (\ref{eq:lowerboundfin}) significantly improves the ability to assess the dimensional scaling of the quantum speed limit.

Furthermore, independent of the number of controls $M=|\{P_{j}\}|$  used to implement the SWAP gate, the unitary operation $V$ given by \eqref{eq:choiceV} is always in the intersection of the stabilizers that correspond to the controls.
Thus, independently of how many and which specific energy levels can be arbitrarily shifted, a SWAP operation cannot be implemented in a time less than $\sqrt{2(N-1)}$.
In order to both obtain the tightest bound \eqref{eq:lowerboundfin} over all appropriate unitaries $V$, and to study the dependence of the bound on the number of the controls $M$, we numerically carry out a maximization over all such $V$.

\subsection{Numerical assessment}
Since the controls are given by a set of projectors $\{P_{j}\}_{j=1}^{M}$, we note that every unitary  $V\in\bigcap_{k}\text{Stab}(iP_{k})$ is of the form 
\begin{align}
V=\left(
\begin{matrix}
\mathcal P &0\\
0 & U
\end{matrix}
\right),
\end{align}
where $\mathcal{P}=\text{diag}(e^{i\theta_{1}},\cdots,e^{i\theta_{M}})$ and $U\in\text{U}(N-M)$. Thus, $V$ can be parameterized by $\frac{(N-M)^{2}}{2}+M$ complex variables.
Once a parametrization is made, the standard optimization algorithms BFGS \cite{BFGS} and L-BFGS-B \cite{LBFGS} are run to maximize the quantity $\frac{\Vert [G,V] \Vert }{\Vert [H_{0},V]\Vert }$ and the best result over all algorithms is taken.
Because the dimension of the parameters scales quadratically with the dimension of the system, we are able to feasibly run up to $N=15$ levels on a desktop computer.

In order to study the tightness of the bound we compare our results with the minimum gate time obtained from numerical gate optimization using GRAPE \cite{Grape, DYNAMOpaper} included in the Python control package QuTip \cite{QuTip1, QuTip2}.
That is, the normalized gate error 
\begin{align}
\label{eq:gateerror}
\epsilon=\frac{1}{\sqrt{2N}}\Vert G-U(T)\Vert, 	
\end{align}
with a goal operation $G$ given by \eqref{eq:Nlevelgoal} is optimized for the N-level system for different gate times $T$ as a function of the number of controls $M$ and the number of levels $N$.
The minimum gate time is found through a binary search over $T$ until the error threshold $\epsilon\leq 10^{-4}$ is reached. \\

\begin{figure}[!h]
\includegraphics[width=1.0\columnwidth]{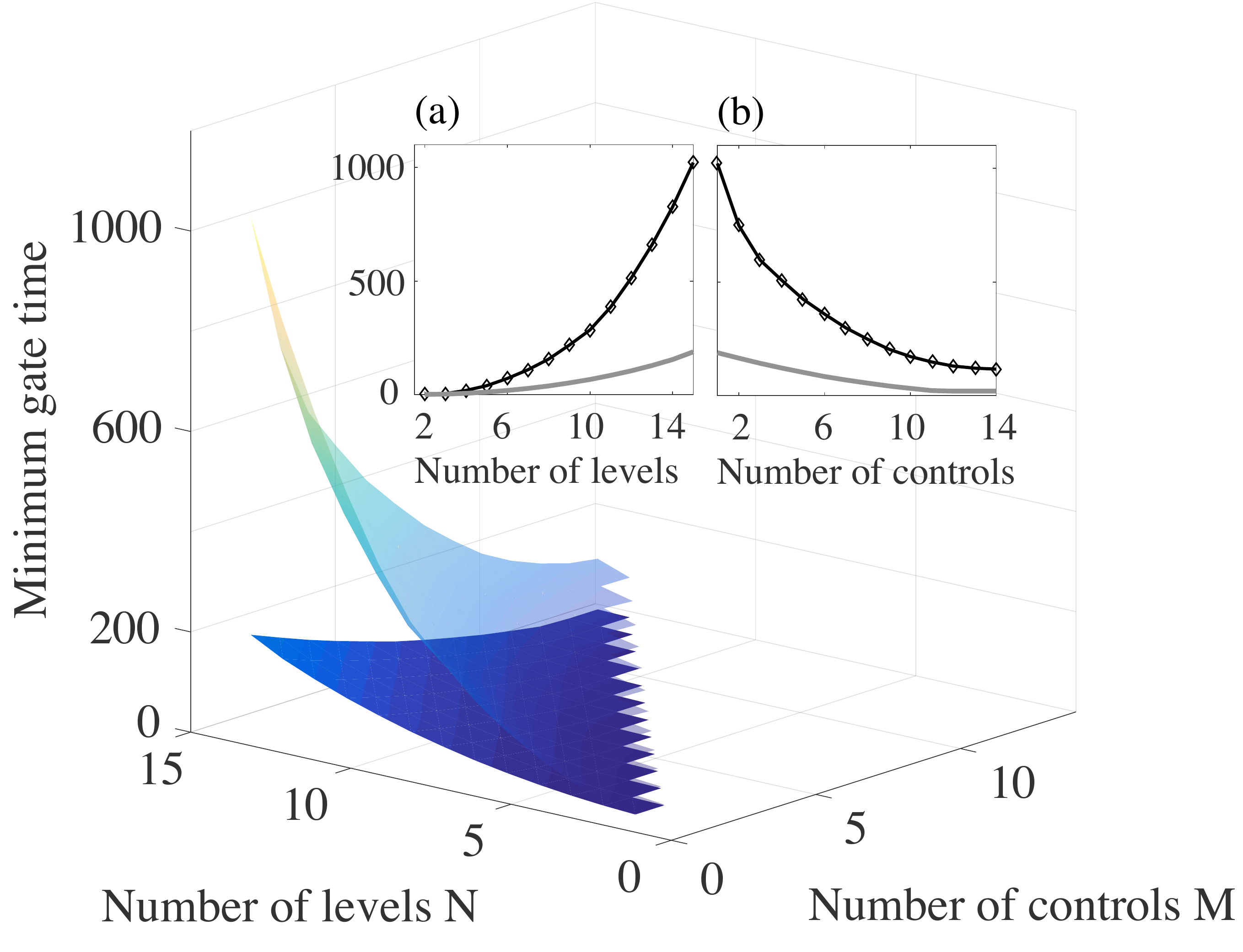}\caption{\label{fig:numericalANA} Comparison of the obtained bound \eqref{eq:lowerboundfin} with the minimum gate time obtained from numerical gate optimization of the gate error \eqref{eq:gateerror} using GRAPE for the N-level system \eqref{eq:nleveldrift} with a SWAP operation \eqref{eq:Nlevelgoal} as the goal. The solid surface plot represents the bound and the transparent surface plot represents the GRAPE data, both as a function of the number of levels $N\in[2,15]$ and the number of controls $M\in[1,14]$. The inset plot show the same data for a fixed number of controls $M=1$ (a), and a fixed number of levels $N=15$ (b), wherein the solid grey line represents the bound and the solid black line the minimum gate time obtained from GRAPE.}
\end{figure}
The results are shown in figure \ref{fig:numericalANA} wherein the solid surface plot represents the numerical evaluation of the bound \eqref{eq:lowerboundfin} as a function of $N\in[2,15]$ and $M\in [1,14]$ and the transparent surface plot represents the minimum gate times obtained from GRAPE. 
The inset plot shows (a) the minimum gate time as function of the number of levels $N$ for a fixed number of controls $M=1$, and (b) the minimum gate time as a function of the number of controls for $N=15$ levels. 
The solid black line represents the data obtained from GRAPE, whereas the solid grey line represents the bound \eqref{eq:lowerboundfin}. 

Although the minimum gate times obtained from GRAPE differ from the numerical optimization of the lower bound \eqref{eq:lowerboundfin}, we first note that the lower bound yields a similar scaling $\propto N^{2}$.
We further note that the constant of proportionality in the case studied is almost exactly 4.
Interestingly, since the plots (a) and (b) are almost symmetric, the numerical analysis suggests that the minimum gate time for implementing a SWAP operation through varying the energy levels of a N-levels system depends in a similar fashion on the dimension of the system and the number of controls being used. 
We can conclude that reducing the number of controls has a similar effect on the minimum gate as increasing the system size.   

\section{Discussion and Conclusion}

We note that the inequality used to obtain the bound holds for \textit{any} unitarily invariant norm, thus giving an additional variable over which to optimize in order to better estimate the minimum time. 
Further work will incorporate optimization over the $\ell_{p}$ norms specifically.
We further note that we have normalized $H_{0}=\frac{h}{\Vert h\Vert}$ using a matrix norm.
In general, if the drift Hamiltonian is normalized by some constant $\mathcal N$, the lower bound \eqref{eq:boundfirst} becomes  
\begin{align}
 T &\geq \frac{\Vert [G,V] \Vert }{\Vert [H_{0},V] \Vert } \mathcal N.
\end{align}
Such a normalization is motivated by physically reasonable constraints, for instance, to ensure that the thermodynamic limit exists or that the total energy density remains finite when the system is scaled up.
We conjecture that is is always possible to find for any dimension a goal operation $G$ and a unitary gate $V$ such that $\frac{\Vert [G,V] \Vert }{\Vert [H_{0},V] \Vert }$ is a dimension independent constant $C$, and that subsequently for such gates $T \geq C\cdot \mathcal N$. Consequently, the quantum speed limit is determined by the physical constraints given by the normalization constant $\mathcal N$.

We have presented a computationally tractable quantum speed limit formula for arbitrary unitary gates which scales with the system dimension.
Furthermore, we have shown that this bound captures physically relevant aspects of the scaling of the minimum time for a specific SWAP gate in an N-level system frequently studied in quantum information science.
We have further shown that the bound can be used to compare the effects of both additional levels, and of additional control fields on the minimum gate time.
One sees that the effects of both are approximately inverse to each other.

We see that although the bound has qualitatively reasonable scaling properties for larger numbers of controls and levels, the true minimum time is no longer well estimated and that the literature still does not contain a systematic methodology for obtaining minimum times in quantum control systems, either analytically or numerically.

Given a control system and a goal unitary, it is an interesting question which control Hamiltonians (perhaps among a restricted available set) will be most beneficial to improving the minimum time.
The form of \eqref{eq:lowerboundfin} can provide some guidance as to the optimal `resource' to add to a quantum control system in order to reduce the minimum time: one should add a control Hamiltonian (from the physically available set) which makes the intersection of the stabilizers of all the control Hamiltonians as small as possible.
Doing so ensures that the set of $V$ over which one maximizes will be as small as possible, and thus give a lesser lower bound on the minimum time.
This direction will form the basis of future work.

\section{Acknowledgements}
The author B.R. acknowledges the NSF (grant CHE-1464569), J.L. the DOE (grant DE-FG02-02ER15344), C.A. the ARO (grant W911NF-16-1-0014), and H.  R. the Templeton Foundation (grant 52265). 

\bibliographystyle{unsrt}
\bibliography{paper1.bib}

\end{document}